\documentstyle[epsfig,here,12pt,twoside,fancyhdr]{article}

\input amssym.def
\input amssym.tex

\def\double{\Bbb}

\def\ccc{{\double C}}

\def\aa{{\cal A}}

\def\hh{{\cal H}}

\def\rr{{\cal R}}

\def\lp{\left(}
\def\rp{\right)}

\def\ul{\underline}

\def\ot{\otimes}

\def\bbb{\begin{equation}}
\def\eee{\end{equation}}
\def\bbbb{\begin{eqnarray}}
\def\eeee{\end{eqnarray}}

\newtheorem{thm}{Theorem}[section]

\newtheorem{cor}{Corollary}[section]

\newcounter{ex}[section]

\def\demo{\noindent\ul{Proof}:\\ \noindent }
\def\edemo{\hfill$\square$\\}

\begin{document}

\title{ Wave-Function renormalization and the  Hopf algebra of Connes and Kreimer}
\author {Florian Girelli $^{(1)}$, Thomas Krajewski $^{(2)}$,   Pierre Martinetti $^{(1)}$  
 \\ \small{\it{ (1) Centre de Physique Th\'eorique, CNRS - Luminy, Case 907, 13288 Marseille Cedex, France}} 
\\  \small {\it{(2) Scuola Internazionale Superiore di Studi Avanzati, Via Beirut 2-4, I-34014, Trieste, Italy  }}}

\maketitle
\abstract{In this talk, we show how the Connes-Kreimer Hopf algebra morphism can be extended when taking into account the wave-function renormalization. This leads us to a semi-direct product of invertible power series by formal diffeomorphisms. }
\newline
\newline
\newline

\section{ The Connes-Kreimer formalism}

As has already been mentioned by A. Connes in his lecture \cite{Connes}, the standard BPHZ recursion formula for the substraction of ultraviolet divergent graphs can be interpreted as an algebraic analogue of the Birkhoff decomposition of loops with values in a infinite dimensional group associated to Feynman diagrams \cite{CK1}. Although the proof of the existence of the Birkhoff decomposition for an infinite dimensional group seems to be a rather difficult task, it is fairly easy to obtain, in the case under consideration,  if one works at the level of the associated algebra of polynomial functions. Besides, it turns out to hold for the group of characters of general graded Hopf algebras.
\cite{kastler}.

\begin{thm}
Let $\hh$ be a graded Hopf algebra, $\aa$ an arbitrary commutative algebra
which is written as a sum (in the sense of vector spaces) of two commutative
sub-algebras $\aa_{-}$ and $\aa_{+}$. Then any character $\gamma$ of $\hh$ with
values in $\aa$ can be decomposed as
\bbb
\gamma=\gamma_{-}^{-1}*\gamma_{+},
\eee
where $\gamma_{\pm}$ is a character of $\hh$ with values in $\aa_{\pm}$.
Moreover,the decomposition is unique up to multiplication on the left by a
character with values in $\aa_{-}\cap\aa_{+}$.
\end{thm}

This is easily applied to the Hopf algebra $\hh$ of Feynman diagrams of a given
field theory (say $\phi^{3}$ in six dimensions). In this case we can choose
$\aa$ to be the algebra of formal Laurent series of finite negative order,
$\aa_{+}$ to be the subalgebra corresponding to positive powers and $\aa_{-}$
the subalgebra of power series with exponent of strictly negative degree.
Evaluation of the diagrams in dimensional regularization yields a character
with values in $\aa$ and its Birkhoff decomposition is nothing but a way to
rewrite the splitting between counterterms and renormalized amplitudes in the
minimal substraction scheme. Note that the convolution product encodes all the
intricacies of the substraction of subdivergences.

In this example, the intersection between the two algebras is empty, so that
the decomposition is unique. However, we could as well include the constants
in $\aa_{-}$, so that the ambiguities of the decomposition are measured by
characters of $\hh$. These characters just associate a number to each Feynman
diagram and their inverses are nothing but arbitrary finite counterterms whose
effects we want to compensate by suitable finite changes of the parameters of
the theory.

Of course, in physically interesting situations we impose normalization conditions to the Green's functions of the theory and it is a  change of the
normalization conditions that yields to  a finite character. Alternatively, we can
use the minimal substraction scheme in dimensional regularization so that we
have a unique solution for the Birkhoff decomposition. However, this theory
has a hidden parameter, the unit of mass $\mu$, and characters for different
values of $\mu$ still differ by a finite counterterm.

A first step involves the associated change of the coupling constant in the massless $\phi^{3}$ theory and has been obtained in \cite{CK2}. Before we give
this result, let us recall that the formal power series in $g$ of the type
\bbb
g+\mathop{\sum}\limits_{n=2}^{\infty}a_{n}g^{n}
\eee
form a group $G_{diff}$ for the usual composition of power series. We shall denote by $G_{diff}^{o}$ the opposite group and by $\hh_{diff}$ the Hopf algebra of
polynomial functions on $G_{diff}^{o}$. The latter is generated as an algebra by
the functions $\alpha_{n}$ with $n>1$, defined on $G_{diff}$ by
\bbb
\alpha_{n}\lp\mathop{\sum}\limits_{k=2}^{\infty}a_{k}g^{k}\rp= a_{n},
\eee
and its Hopf algebra structure is derived from the group law of $G_{diff}$.

\begin{thm}
Let $Z_{1}(g)$ and $Z_{3}(g)$ be the formal of power series in the coupling
constant $g$ that correspond to sum of the 1PI diagrams of the 2 and 3 point
functions. Then the expansion
\bbb
gZ_{1}(g)Z_{3}^{-3/2}(g)=g+\mathop{\sum}\limits_{n=2}^{\infty}z _{n}g^{n}
\eee
generates a Hopf algebra homomorphism $\Psi(\alpha_{n})=z _{n}$ from
$\hh_{diff}$ to $\hh$.
\end{thm}

Using the coproduct of $\hh$ one can define a right action of $G$ on $\hh$ by
\bbb
f^{\gamma}=(\gamma\ot id)o\Delta
\eee
for any $\gamma\in G$ and $f\in\hh$. This is an action by algebra
homomorphisms and it extends to the algebra of formal power series with
coefficients in $\hh$.

On the other side, the Hopf algebra morphism $\Psi$ induces at the level of
characters a group morphism from $G$ to $G_{diff}$. Since any element of
$G_{diff}$ acts on formal power series by composition, we have another action
of $G$ on formal power series with coefficients in $\hh$. The combination of
the two actions leave the coupling constant invariant in the following sense.

\begin{cor}
Let $\gamma\in G$ and denote by $\Psi_{\gamma}$ the associated formal
diffeomorphism. Then
\bbb
\lp \psi_{\gamma}(g)^{-1}Z_{g}(\psi_{\gamma}(g)^{-1})\rp^{\gamma}=gZ_{g}(g).
\eee
\end{cor}

When evaluated on the character $\gamma_{+}$, this is just the statement that
the renormalized theory is invariant provided the effect of the finite
counterterm $\gamma$ has been compensated by trading the old coupling constant
$g$ for the new one $\psi_{\gamma}(g)$.

To complete the picture of reparametrization invariance, we have to study the
effect of the combined change of counterterm and coupling constant on the
Green's functions. This will be achieved by extending the previous morphism to
take into account finite wave function renormalization.
\section{Wave-function renormalization}
Suppose that we consider two renormalization schemes $\rr$ and $\rr'$ and
denote by primed indices the renormalization factors associated to the scheme
$\rr'$. By this we simply mean that we have two ways to compute the renormalized Feynman diagrams that differ only by finite values. Then the new coupling constant $g'$ is determined as a function of the old one by requiring that the bare coupling constants are the same for both schemes
\bbb
g_{0}'(g'(g))=g_{0}(g)
\eee
which yields directly to the definition of the previous morphism.

As is well known, the Green's functions are not invariant under this transformation since it involve also a change of the normalization of the field. Accordingly, the $N$-point functions in the two renormalization schemes are related by
\bbb
\Gamma_{N}^{'}(p,g')=\zeta(g)\Gamma_{N}(p,g)
\eee
with
\bbb
\zeta(g,\rr',\rr)=\frac{Z_{3}^{\rr'}}{Z_{3}^{\rr}}.
\eee
Note that physics of the system (i.e. the S-matrix) is invariant by such a transformation.

We need now to extend $\hh _{Diff}$ to take account of the effect of wave-function renormalization, in the spirit of what has been done by Connes and Kreimer.

First let $G_{pow}$ be the group (for usual multiplication) of power series of the type $1 + \sum_{n \geq 1} c_n g^n$ with $c_n \in \ccc$.
\\ Define $\tilde G= G_{diff} \ltimes G_{pow}$ with the group law
\bbb
(S, T). (S', T')= (SoS', ToS' T') \; \; \; \forall S, S' \in G_{diff}, T, T' \in G_{pow}.
\eee

Let us denote by $\gamma$ the character of $\hh$ associated to the change from $\rr$ to $\rr^{'}$. Motivated by the previous considerations on wave function renormalization, we define the following function on $G$:
\bbb
\zeta _{\gamma}(x) = \frac{Z_3 ^{\gamma}[ \psi _{\gamma}^{-1} (x)]}{Z_3 (x)}.
\eee
It describes wave function renormalization and leads to the following result.

\begin{thm}
The application  $\gamma \rightarrow ( \psi _{\gamma} ^{-1}(x), \zeta _{\gamma}(x))$ is a group morphism from $ G$ to  $ G_{diff} \ltimes G_{pow}$.
\end{thm}
\demo
We  prove this in two steps. First we show that the application $\zeta _{\gamma}(x)$ is well defined because it must be a formal power series with scalar coefficients whereas its definition involves $\hh$ valued coefficients. Then we prove that the group law is satisfied.

By construction we have
\bbb
\psi _{\gamma} ^{-1} (x)Z_1^{\gamma}( \psi _{\gamma} ^{-1})(Z_3^{-3/2})^{\gamma}( \psi _{\gamma} ^{-1}) =  x Z_1(x)Z_3^{-3/2}(x).
\eee
So by reordering, we get
\bbb
\frac{(Z_3)^{\gamma}( \psi _{\gamma} ^{-1})}{Z_3(x)}= \left(\frac{ \psi _{\gamma} ^{-1}Z_1  \psi _{\gamma} ^{-1} )}{x Z_1 (x)}\right)^{2/3}.
\eee
This just asserts that a diagram with 2 external lines has to be equal to a diagram with 3 external lines, the only possibility is that it is a scalar. So $\zeta _{\gamma}(x) $ is well defined.

Moreover, it is invariant under the action of $G$, so that
\begin{eqnarray}
\zeta _{\gamma \gamma'}(x)&=& \frac{Z_3 ^{\gamma\gamma'}[ \psi _{\gamma\gamma'} ^{-1} (x)]}{Z_3 (x)}\\
&=& \left[\frac{ {Z_3 ^{\gamma} ( \psi _{\gamma} ^{-1} (x )o \psi _{\gamma'} ^{-1} (x))}}{{Z_3 ^{\gamma}( \psi _{\gamma'} ^{-1} (x))}}\right] ^{\gamma'} \frac{Z_3 ^{\gamma '}( \psi _{\gamma'} ^{-1} (x))}{Z_3(x)} \\
&=&  \zeta _{\gamma}( \psi _{\gamma'} ^{-1} (x))  \zeta _{\gamma '} (x)
\end{eqnarray}
which is just the product law.
\edemo

Accordingly, after evaluation of $Z_1(x)$ and $Z_3(x)$ on a character, we get the transformation law of the 2- and 3-points functions by the following rules:
\begin{eqnarray}
Z_1 ^{\gamma}( \psi _{\gamma} ^{-1} (x))=\zeta _{\gamma}^{3/2}(x) Z_1 (x)\\
Z_3 ^{\gamma}( \psi _{\gamma} ^{-1} (x))=\zeta _{\gamma}(x) Z_3 (x).
\end{eqnarray}
Note that $ \zeta _{\gamma}(x) $ is independent of any regulation procedure: it is a purely diagrammatical object.

\section{Concluding remarks}
We have shown that the renormalization of the wave-function yields also to a morphism from $G$ to $ \tilde G= G_{diff} \ltimes G_{pow}$ which is purely diagrammatical.

We should also point out that this is a very particular case of a general statement about the reparametrization invariance of the functional integral that is currently under investigation using Hopf algebraic techniques. For instance, the previous morphsim can be obtained through a two variable power series as follows: let us denote by $Z_{1,2i}$ (resp. $Z_{3,2i}$) the sum of all 1PI diagrams with $i$ loops, weighted with their symmetry factor, that contribute to the 3-point function (resp. the 2-point function). Then
\begin{eqnarray}
X(x,y)&=& x(1+Z_{1,2}x^2y^3+Z_{1,4}x^4y^6+\dots+Z_{1,2i}x^{2i}y^{3i}+\dots )\\
Y(x,y)&=& y(1-Z_{3,2}x^2y^3-Z_{3,4}x^4y^6+\dots-Z_{3,2i}x^{2i}y^{3i}+\dots )^{-1}
\end{eqnarray}
defines a morphsim of two variables that preserves the foliation of the first quadrant by the curves $x^2y^3$ from which we recover the coupling constant renormalization (change of the curve) and wave function (modification of the coordinate on the curve).

\end{document}